\documentclass[10pt, conference, letterpaper]{ IEEEtran}
\IEEEoverridecommandlockouts
\usepackage{cite}
\usepackage{amsmath,amssymb,amsfonts}
\usepackage{algorithmic}
\usepackage{graphicx}
\usepackage{textcomp}
\usepackage{xcolor}
\usepackage{setspace}
\usepackage{enumitem}
\usepackage{hyperref}
\usepackage{xurl}
\usepackage{cleveref}
\usepackage{array}
\usepackage{multirow}

\usepackage{booktabs}
\usepackage{tabularray}

\def\BibTeX{{\rm B\kern-.05em{\sc i\kern-.025em b}\kern-.08em
    T\kern-.1667em\lower.7ex\hbox{E}\kern-.125emX}}
\begin{document}

\title{Decompose to Understand, Fuse to Detect: Frequency-Decoupled Anomaly Detection for Encrypted Network Traffic}


\author{\IEEEauthorblockN{Xinglin Lian\IEEEauthorrefmark{2}, Chengtai Cao\IEEEauthorrefmark{3}, Ting Zhong\IEEEauthorrefmark{2}, Yong Wang\IEEEauthorrefmark{4}, Kai Chen\IEEEauthorrefmark{4}, Fan Zhou\IEEEauthorrefmark{2}\IEEEauthorrefmark{5}\IEEEauthorrefmark{1}}
\IEEEauthorblockA{\IEEEauthorrefmark{2} School of Information and Software Engineering, University of Electronic Science and Technology of China}
\IEEEauthorblockA{\IEEEauthorrefmark{3} Department of Computer Science, City University of Hong Kong}
\IEEEauthorblockA{\IEEEauthorrefmark{4} Department of Computer Science and Engineering, Hong Kong University of Science and Technology}
\IEEEauthorblockA{\IEEEauthorrefmark{5} Key Laboratory of Intelligent Digital Media Technology of Sichuan Province}
\IEEEauthorblockA{\IEEEauthorrefmark{1} Corresponding Author: fan.zhou@uestc.edu.cn}
}

%

\maketitle

\begin{abstract}
Network traffic anomaly detection represents a critical cybersecurity task, yet widespread encryption makes this task increasingly challenging. In response, image-based methods that model traffic as visual patterns have emerged as the dominant approach. However, this work pioneers the identification of a pervasive ``full-frequency'' characteristic and an associated limitation termed ``spectral mismatch'' within this paradigm. Specifically, while encrypted traffic exhibits prominent high-frequency components, mainstream reconstruction methods demonstrate an inherent bias toward learning low-frequency information. This fundamental mismatch results in incomplete representations that consequently degrade anomaly detection performance. To address this challenge, we propose FreeUp, a novel frequency-decoupled framework designed explicitly for encrypted traffic analysis. FreeUp decomposes traffic data into distinct low- and high-frequency bands, processing them through separate, dedicated branches along with a customized training strategy that ensures stable and independent frequency-specific learning. Furthermore, recognizing that simple reconstruction error proves inadequate for evaluating dual-branch architectures, we introduce an uncertainty-inspired fusion scoring mechanism. This mechanism quantifies the reconstruction uncertainty of the frequency-specific branches and dynamically integrates their outputs, yielding a more comprehensive and reliable anomaly score. Extensive experiments across multiple benchmarks demonstrate that FreeUp consistently outperforms state-of-the-art baselines. The code is available at \url{https://github.com/ikun0124/FreeUp}.
\end{abstract}


\section{Introduction}
\label{sec:Introduction}

Network traffic anomaly detection is a cornerstone of modern cybersecurity, tasked with identifying unauthorized, malicious, or otherwise covert traffic patterns that threaten network integrity~\cite{zeng2024leveraging}. Its effective implementation not only safeguards sensitive data assets~\cite{li2023bijack, zheng2024multi, zeng2025enhancing} but also ensures communication service quality for end users~\cite{zhang2025continual, han2024ecnet, zeng2024fga}. However, this task is becoming increasingly complex. The exponential growth in traffic volume and the continuous evolution of sophisticated attack techniques pose significant challenges for precise anomaly detection~\cite{zhong2023dynamic}. Compounding these difficulties, the widespread adoption of encryption, driven by escalating concerns over user privacy, introduces a new obstacle~\cite{zhong2025noise}. While essential for protecting sensitive information, encryption obscures packet payload features that are crucial for distinguishing between normal and malicious traffic. Consequently, traditional detection methods, such as those reliant on character matching algorithms, are largely rendered ineffective in modern encrypted environments~\cite{zhang2024aoc}.

Deep learning approaches, especially those based on a zero-positive learning paradigm, have emerged as a dominant solution for network traffic anomaly detection~\cite{lian2025facing}. These methods operate on the principle of exclusively learning normal traffic distributions and flagging significant deviations as anomalies. In practice, a widely adopted strategy is to transform raw network traffic into 2D image representations, which preserves payload information and enables the application of powerful computer vision techniques~\cite{zhao2024novel, liu2024atvitsc, lian2025mfr}. Inspired by successes in image analysis, most of these approaches are reconstruction-based~\cite{lunardi2023arcade, lian2025mfr, akcay2018ganomaly}. During training, a model learns to recreate images of normal traffic with high fidelity. In the inference phase, anomaly scores are derived from reconstruction errors: low errors indicate normalcy, while high ones imply potential threats. As a result, reconstruction quality becomes the critical factor in anomaly detection~\cite{zheng2023semi}.

\begin{figure}[t]
\centerline{\includegraphics{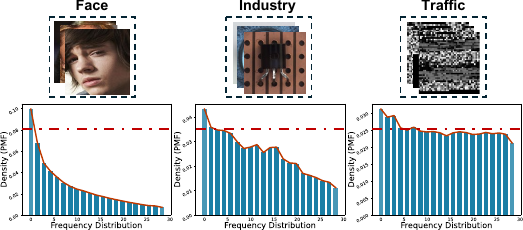}}
\vspace{-0.18cm}
\caption{Illustration of the full-frequency phenomenon via frequency distribution statistics across natural and traffic images.}
\label{fig:decoupling1}
\end{figure}

Despite notable progress, a fundamental limitation persists, stemming from the unique characteristics of traffic images. Unlike natural images that are rich in semantic content, traffic images often contain limited textual or semantic information due to encryption~\cite{zheng2023semi, lin2022bert}, as illustrated in~\figurename~\ref{fig:decoupling1}. Our investigation reveals this distinction extends to the frequency domain. Fourier Transform analysis confirms a key spectral distinction: while natural images are dominated by \textbf{low-frequency} components, encrypted traffic images exhibit a unique \textbf{full-frequency} phenomenon, preserving significant high-frequency components. This is where the core problem arises. Reconstruction models have an inherent \textbf{spectral bias}; they excel at learning low-frequency features but consistently struggle to capture high-frequency variations~\cite{fridovich2022Spectral, chen2024frequency}. This creates a fundamental \textbf{spectral mismatch}: the models' low-frequency preference is well-suited for reconstructing natural images; however, it is fundamentally at odds with the full-frequency nature of traffic image data. Consequently, while these models reconstruct natural images with high fidelity~\cite{yao2024glad}, they produce poor reconstructions for traffic images. The incomplete pattern understanding impedes their ability to discriminate between normal and anomalous traffic, leading to unreliable identification. This spectral mismatch, therefore, poses the core challenge for accurate anomaly detection in full-frequency encrypted traffic data.

Motivated by the above insights, we propose \textbf{FreeUp}, a novel \underline{\textbf{Fre}}qu\underline{\textbf{e}}ncy-deco\underline{\textbf{up}}led framework, which aims to free up frequency components in encrypted traffic by mitigating spectral mismatch in representation learning. Our FreeUp decomposes each traffic image into complementary low- and high-frequency bands and processes them via independent branches. Each frequency-constrained autoencoder specializes in modeling its assigned frequency spectrum. A key aspect of our design is the final reconstruction step: the output of one frequency branch is integrated with the original input of its complementary band. This design, combined with frequency decoupling, offers two significant advantages. \textbf{(i) Focused Learning}: By tasking each branch with modeling only its assigned frequency component, we significantly alleviate the complexity of modeling full-frequency patterns in one pass; \textbf{(ii) Enhanced Reconstruction Stability}: The unique integration strategy provides a stable learning signal for each branch. For instance, the low-frequency branch's reconstruction quality is assessed against a ``perfect'' high-frequency ground truth (the original input), making its training stable. This ensures that the model's ability to learn one set of patterns is not compromised by difficulties in learning the other. The two advantages jointly alleviate the spectral bias issue common in traditional models, which prefer low-frequency content but neglect high-frequency information.

Having established the FreeUp framework for high-fidelity reconstruction, the next critical step is translating reconstruction quality into a sensitive anomaly score. Traditional methods often rely on \textbf{scalar reconstruction errors} to quantify anomaly degree, which overlook valuable distributional information. To overcome this, we adopt a more nuanced approach: a lightweight evidential learning method is employed to independently model the \textbf{reconstruction uncertainty distribution} associated with each frequency branch. This aligns perfectly with our decoupled design, allowing us to pinpoint anomalies that may only manifest in a specific frequency band. Anomalous samples, deviating from the learned normal distribution, will thus yield higher uncertainty scores in their respective branches~\cite{lian2025facing}. However, relying solely on independent scores is insufficient, as some complex anomalies reveal themselves through subtle deviations in multiple frequencies. Therefore, we propose an uncertainty-inspired dynamic fusion strategy to create a holistic final judgment. This strategy integrates the detection results by combining evidential distribution parameters from multiple branches/views into a joint distribution. Guided by a multi-task training objective, the fused distribution dynamically assigns weights to each branch’s contribution, enabling the model to adaptively capture uncertainty across both low- and high-frequency views. As a result, the uncertainty derived from this adaptively fused distribution serves as the final anomaly indicator, enabling a more comprehensive and accurate anomaly score. We validated our framework on multiple real-world encrypted traffic datasets, where the experimental results consistently demonstrate significant improvements over state-of-the-art baseline methods.

In summary, our key contributions are fourfold:
\begin{itemize}
    \item We are the first to systematically identify and characterize the spectral mismatch challenge between the full-frequency nature of encrypted traffic and the inherent low-frequency bias of reconstruction-based models.
    \item We propose FreeUp, a novel frequency-decoupled framework designed explicitly to address this fundamental challenge by enabling dedicated representation learning for each frequency band.
    \item We design an uncertainty-inspired dynamic fusion mechanism that adaptively integrates multi-view frequency uncertainty, yielding comprehensive anomaly evaluation.
    \item  Through extensive experiments on multiple benchmark datasets, we demonstrate that FreeUp significantly outperforms state-of-the-art methods.
\end{itemize}
\section{Motivation}\label{sec:Motivation}
\subsection{Full Frequency Phenomenon}

The central premise of our work is that encrypted traffic images exhibit spectral properties distinct from the natural images commonly studied in anomaly detection. To empirically ground this motivation, we conduct a systematic, comparative analysis of their frequency-domain characteristics.

Our investigation compares encrypted traffic images with two representative natural image categories often found in anomaly detection: the face spoofing dataset CelebA~\cite{liu2015deep} and the industrial defects dataset MVTec~\cite{bergmann2019mvtec}. For fair comparison, we randomly sampled 1,000 images from each category and resized them to match the resolution of traffic images. We then applied a 2D Fourier Transform to each image to compute its power spectra. As shown in~\figurename~\ref{fig:decoupling1}, the averaged power spectra reveal a clear distinction across categories.

\noindent \textbf{Low-Frequency Dominance in Natural Images:} Our investigation of natural images, including both facial photographs and industrial items, reveals a consistent spectral pattern: their energy is overwhelmingly concentrated in the low-frequency domain, with a sharp drop-off in the high-frequency range. This is a well-established principle in computer vision, where dominant low-frequency components correspond to an image's semantic structures and outlines, such as object backgrounds and shapes~\cite{chen2024frequency}. Conversely, high-frequency components encode fine details, textures, and edges, serving a complementary role in refining the base structures~\cite{tu2024hyperspectral}. Given this abundance of low-frequency content and scarcity of high-frequency information, we term this characteristic the ``low-frequency dominance'' phenomenon. This spectral composition results in a spatial pattern that is coherent and semantically interpretable, a hallmark of real-world imagery.

\noindent \textbf{Full-Frequency Nature of Encrypted Traffic:} The encryption process fundamentally alters traffic data, disrupting the original byte-level semantics and creating a visually disordered, noise-like appearance when converted into images~\cite{lian2025mfr, liu2024atvitsc}. This spatial randomness has a distinct and revealing signature in the frequency domain. As shown in~\figurename~\ref{fig:decoupling1}, our spectral analysis reveals that the high-frequency components of traffic images are exceptionally prominent, often exhibiting magnitudes equivalent to their low-frequency counterparts. This abundance of high-frequency power corresponds to the abrupt, pixel-to-pixel variations inherent in the encrypted data. As a result, any potential low-frequency structures are effectively overwhelmed, obscuring coherent and interpretable traffic pattern information. While some works have partially noted these spatial characteristics~\cite{lian2025mfr, farrukh2023senet}, their underlying spectral properties and conflict with model biases remain underexplored. We, therefore, present the first formal characterization of this phenomenon. Specifically, we define this unique spectral distribution as ``full-frequency'': a state where both low- and high-frequency components are highly active and coexist with comparable intensity, posing a significant challenge for effective representation learning.

\subsection{Motivation: The Inherent Conflict}
\noindent\textbf{Spectral Mismatch Limitation:} A well-documented phenomenon in deep learning is ``spectral bias'', where models, particularly autoencoders, tend to favor low-frequency patterns while struggling to capture high-frequency details~\cite{chen2024frequency, fridovich2022Spectral}. For natural image analysis, this bias is often harmless or even advantageous, because the primary semantic content (e.g., object shapes) is concentrated in low frequencies, allowing models to perform well despite ignoring some fine textures~\cite{yao2024glad}. However, this bias becomes a critical vulnerability in the context of ``full-frequency'' traffic images. When a standard reconstruction model is trained on such images, its inherent bias leads the model to capture the low-frequency components effectively while largely ignoring the equally important high-frequency ones. This results in an incomplete and distorted understanding of the normal data distribution. Consequently, the reliability of anomaly scoring is severely compromised, as the model lacks the high-frequency fidelity needed to detect subtle deviations. We formulate this representational learning conflict as ``spectral mismatch''.

\begin{figure}[t]
\centerline{\includegraphics{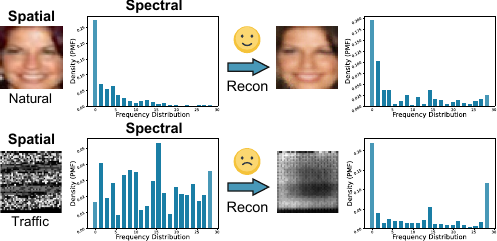}}
\vspace{-0.18cm}
\caption{Spectral mismatch in traffic images: limitations of the full-frequency phenomenon caused by the ``spectral bias'' issue.}
\label{fig:decoupling2}
\end{figure}

\noindent\textbf{Empirical Validation:} To empirically validate this conflict, we train a state-of-the-art network traffic anomaly detection model, MFR~\cite{lian2025mfr}, on two datasets: face spoofing dataset CelebA~\cite{liu2015deep} and encrypted traffic dataset DoHBrw2020~\cite{montazerishatoori2020detection}. \figurename~\ref{fig:decoupling2} clearly illustrates the impact of ``spectral mismatch''. Due to their ``low-frequency dominance'' nature, face spoofing images are reconstructed with high spatial fidelity, and their spectral distributions remain largely intact. In sharp contrast, the reconstructed traffic images experience substantial quality degradation, preserving only coarse-grained brightness profiles. The corresponding power spectra confirm the cause: the model fails to retain high-frequency information, severely disrupting the original spectral distribution between low- and high-frequency components. This experiment offers empirical evidence of the spectral mismatch and directly motivates the need to overcome this fundamental limitation.

\noindent\textbf{Motivation:} The systemic failure in reconstruction poses a critical consequence for anomaly detection. When a model produces similarly low-quality reconstructions for both normal and anomalous traffic, the resulting anomaly scores become indistinguishable. This renders the detector unreliable, especially in challenging zero-positive scenarios~\cite{lian2025facing}. Motivated by this challenge, we propose a new paradigm: decoupling the learning process by decomposing the ``full-frequency'' signal into multi-view, i.e., low- and high-frequency bands. In this framework, each branch is dedicated to modeling only its assigned portion of the spectrum. This targeted strategy directly counters spectral bias by preventing the model from defaulting to the easily learned low-frequency patterns. By facilitating a better full-spectrum understanding of traffic data, our method constructs a more complete representation of normal patterns, enhancing the distinguishability of anomalous samples with distributional deviations.
\begin{figure*}[t]
\centerline{\includegraphics{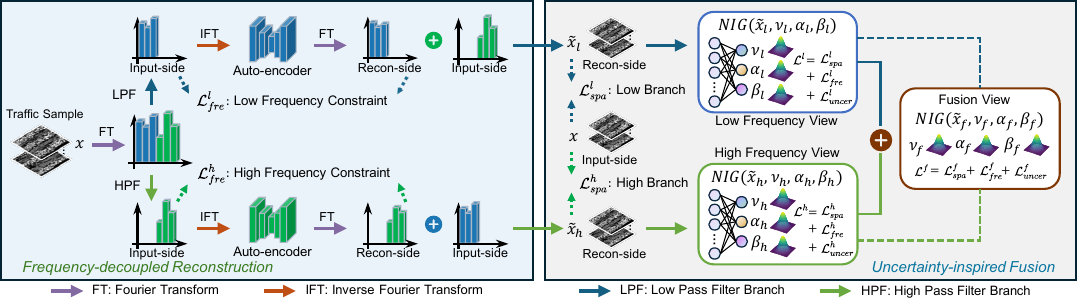}}
\vspace{-0.18cm}
\caption{The overall framework of FreeUp. (1) The model first decouples the input into low- and high-frequency bands, which are reconstructed by separate networks to mitigate spectral bias. (2) An uncertainty-inspired fusion pipeline then adaptively integrates evidential uncertainty from multi-view branches.}
\label{fig:Framework}
\end{figure*}
\section{Preliminaries}
This work investigates the zero-positive anomaly detection problem in the context of network traffic analysis. Let $\mathcal{X} = \{\boldsymbol{x}_{1}, \boldsymbol{x}_{2},...,\boldsymbol{x}_{N}\}$ denote a set of $N$ normal samples; the objective of detection models is to learn the distributional patterns of normal samples during training. For inference, the model assigns an anomaly score to each test sample $\boldsymbol{x}_{\text{test}} \in \mathcal{X}_{\text{test}}$. Higher scores indicate higher anomaly likelihood.

Following established methods~\cite{lian2025facing, lian2025mfr}, we convert raw traffic data into an image-like format. Each traffic flow, $\boldsymbol{x}_{i}$, is treated as a sequence of packets. The bytes of each packet are reshaped into an $H \times W$ image, and these images are then concatenated chronologically. This process yields a final tensor for each sample, $\boldsymbol{x}_{i} \in \mathbb{R}^{P \times H \times W}$, where $P$ is the number of packets, and $H$ and $W$ are spatial dimensions.

\section{Methodology}
\label{sec:Methods}
This section presents FreeUp, which frees up network traffic anomaly detection from ``spectral mismatch''. As shown in~\figurename~\ref{fig:Framework}, our framework comprises two main components: frequency-decoupled reconstruction and uncertainty-inspired fusion. First, to mitigate the ``spectral mismatch'', we decompose the input into low- and high-frequency bands and train specialized reconstruction networks. This enables more complete learning of the normal patterns. Second, we employ a parameterized evidential learning to quantify reconstruction uncertainty in a multi-view manner. Anomalous samples, which are reconstructed poorly, will yield higher uncertainty. Finally, these multi-view frequency-specific uncertainties are dynamically fused to generate a unified anomaly score.

\subsection{Frequency-decoupled Reconstruction}
\noindent\textbf{Frequency Decoupling:} As discussed in Section~\ref{sec:Motivation}, the inherent ``full-frequency'' characteristics of traffic images conflict with the low-frequency bias of standard reconstruction models. This conflict triggers a ``spectral mismatch'', which limits the model’s capacity to learn a complete representation. To mitigate this, we introduce an explicit frequency decoupling mechanism in the learning process. We decompose the input traffic image into two distinct views: a low-frequency image containing the general structure and a high-frequency image capturing the fine-grained details.
This separation forces the model to learn from dedicated branches, each branch specializing in a frequency-specific range. By decomposing the learning tasks, we mitigate frequency crosstalk and reduce the ``full-frequency'' modeling complexity. Consequently, each frequency branch can achieve a more accurate reconstruction and complete information representation.

As shown in~\figurename~\ref{fig:Framework}, we first apply the Fourier Transform $\mathcal F(\cdot)$ to convert traffic sample $\boldsymbol x$ into frequency domain:
\begin{equation}
{\mathcal F(:, u, v)}=\sum_{h=0}^{H-1} \sum_{w=0}^{W-1} \boldsymbol{x}(:, h, w) \mathrm{e}^{-j 2 \pi\left(\frac{u h}{H}+\frac{v w}{W}\right)},
\label{eq:eq1}
\end{equation}
where $j$ denotes the imaginary unit, and $(h,w)$ and  $(u,v)$ represent spatial and frequency indices, respectively.

To decompose different spectral components, we apply Gaussian low- and high-pass filters in the frequency domain:
\begin{equation}
\mathcal F^{'}_{lpf}(:,u, v)=\mathcal F(:,u, v) \cdot \mathcal H_{lpf}(D),
\label{eq:eq2}
\end{equation}
\begin{equation}
\mathcal F^{'}_{hpf}(:,u, v)=\mathcal F(:,u, v) \cdot \mathcal H_{hpf}(D),
\label{eq:eq3}
\end{equation}
where $\mathcal H_{lpf}(D)$ and $\mathcal H_{hpf}(D)$ are standard Gaussian filters with cutoff threshold $D$, effectively controlling the separation of low- and high-frequency bands. In addition, $\mathcal F^{'}(\cdot)$ denotes the filtered spectrum.

Next, we transform the filtered spectra back into the spatial domain using the inverse Fourier Transform $\mathcal{F}^{-1}(\cdot)$:
\begin{equation}
\boldsymbol x_{lpf}=\mathcal{F}^{-1}(\mathcal F^{'}_{lpf}(:, u, v)), \boldsymbol x_{hpf}=\mathcal{F}^{-1}(\mathcal F^{'}_{hpf}(:, u, v)),
\label{eq:eq4}
\end{equation}
\begin{equation}
\mathcal F{^{-1}(:,h, w)}=\sum_{u=0}^{H-1} \sum_{v=0}^{W-1} \mathcal F{^{'}(:, u, v)} \mathrm{e}^{j 2 \pi\left(\frac{u h}{H}+\frac{v w}{W}\right)}.
\label{eq:eq5}
\end{equation}

This frequency decomposition yields two spatial-domain representations: the low-frequency branch $\boldsymbol x_{lpf}$ and the high-frequency branch $\boldsymbol x_{hpf}$. These decoupled views are subsequently used for frequency-specific modeling.

\noindent\textbf{Frequency-constrained Auto-encoder:} The decomposed traffic image ${\boldsymbol x_{lpf}}$ and ${\boldsymbol x_{hpf}}$ are then independently fed into two autoencoder models for frequency-specific learning. Each autoencoder is dedicated to reconstructing its frequency-specific, without interference from the other component. This explicit separation naturally enforces the model to focus on frequency-specific distributions and effectively mitigates the reconstruction bias toward low-frequency content. To achieve this, we employ an autoencoder architecture enhanced with channel-spatial attention~\cite{lian2025mfr}. This mechanism helps the model emphasize the most informative spatial and channel-wise regions in the feature maps. Hence, it enhances the frequency-specific representation quality. To ensure computational efficiency, we adopt a concise encoder–decoder architecture for each branch.

As illustrated in the left part of~\figurename~\ref{fig:Framework}, we further introduce a frequency-complement integration mechanism. In this design, each reconstruction-side frequency branch integrates the complementary frequency component from the input-side as auxiliary information. This integration provides a more complete frequency basis and helps stabilize the reconstruction process. The complementary frequency serves as a reliable prior, helping the model avoid degradation in cases of difficult reconstruction. By cross-branch integration, we obtain two view frequency-specific representations: low- $\widetilde{\boldsymbol x}_{l}$ and high-frequency $\widetilde{\boldsymbol x}_{h}$, respectively. This process is:
\begin{equation}
    \boldsymbol{\widetilde{x}}_{l} =  \mathcal F^{-1}({\mathcal F(\operatorname{AE}(\boldsymbol{x}_{lpf}})) + {\mathcal F(\boldsymbol x_{hpf}})),
\label{eq:eq6}
\end{equation}
\begin{equation}
    \boldsymbol{\widetilde{x}}_{h} = \mathcal F^{-1}({\mathcal F(\operatorname{AE}(\boldsymbol{x}_{hpf}})) + \mathcal F(\boldsymbol x_{lpf})).
\label{eq:eq7}
\end{equation}
During the reconstruction phase, our objective is to accurately reconstruct the multi-view frequency-specific branches. Existing works commonly rely on a spatial-domain Mean Squared Error (MSE) to realize reconstruction~\cite{zhang2024aoc, lian2025mfr, yang2025unflows}. However, this only captures the spatial nature but cannot directly address the unique ``full-frequency'' nature of traffic data. Considering this, we propose a frequency-constrained strategy that enforces frequency consistency during reconstruction. Inspired by prior work on spectral supervision~\cite{lin2023catch}, our proposed mechanism aims to: (1) Suppress frequency leakage by minimizing the presence of high-frequency content in low-frequency reconstructions, and vice versa. (2) Enhance spectral fidelity by applying an MSE loss in the frequency domain, computed on the Fourier-transformed output. Thus, we compute reconstruction losses in both spatial and frequency domains. The total reconstruction loss is:
\begin{equation}
\mathcal L_\mathrm {REC}= \underbrace{\Vert {\boldsymbol {x}} - \boldsymbol{\widetilde{x}}\Vert_1}_{\text{Spatial Reconstruction}} + \underbrace{\Vert{ {\mathcal F^{'}_{{\boldsymbol {x}}}(u,v)}} - {\mathcal F^{'}_{\boldsymbol{\widetilde{x}}}(u,v)}\Vert_1 }_{\text{Frequency Reconstruction}}\cdot\mathcal{\lambda}_{\mathrm{F}} .
\label{eq:eq8}
\end{equation}
Here, for clarity, we use $\boldsymbol{\widetilde{x}}$ to denote both $\widetilde{\boldsymbol x}_{l}$ and $\widetilde{\boldsymbol x}_{h}$, representing the two branches, and $\mathcal F^{'}(\cdot)$ denotes the filtered spectrum in Equations~\ref{eq:eq2} and~\ref{eq:eq3}. $\mathcal{\lambda}_{\mathrm{F}}$ is a hyperparameter to control the contribution of the frequency-constrained loss.

\subsection{Uncertainty-inspired Fusion} 
\noindent\textbf{Uncertainty-based Anomaly Quantification:} Traditional anomaly detection methods rely on scalar reconstruction errors to quantify anomaly degree, which does not directly exploit the intrinsic distributional differences. Inspired by the inter-sample difference between normal and anomalous data, we introduce an effective and lightweight uncertainty quantification strategy to serve as the basis for anomaly scoring~\cite{lian2025facing}. Epistemic uncertainty represents the model’s confidence in its reconstruction output. During training, the model is exposed only to normal samples, leading to high confidence (i.e., low uncertainty). In contrast, anomalous samples, unseen during training, typically yield low confidence (i.e., high uncertainty). This discrepancy in uncertainty distribution provides a natural metric for anomaly detection. Unlike traditional reconstruction-based approaches, our method directly leverages distributional deviations in uncertainty to evaluate anomalies, offering a more principled detection mechanism~\cite{lian2025facing}.

To achieve this, we use an evidential learning framework to efficiently quantify uncertainty through higher-order prediction distributions. We first consider traffic data $\boldsymbol {x}$ that follows a Gaussian distribution, which is frequently defined during standard analysis scenarios~\cite{fu2023point, lian2025facing}. Assuming the reconstruct $\widetilde{\boldsymbol x}_{l}$ and $\widetilde{\boldsymbol x}_{h}$ are trained well and similar $\boldsymbol {x}$, and let its distribution be modeled using the Normal Inverse-Gamma (NIG) distribution, parameterized by $(\mu, \sigma^2)$ $\sim$ NIG(${v,\alpha,\beta}$). This allows estimation of both the mean $\mu$ and variance $\sigma^2$ of the reconstruction. The reconstruction uncertainty (used as the anomaly score) is quantified as the variance of the mean:
\begin{equation}
\mathrm{Var}[\mu]=\frac\beta{v(\alpha-1)}.
\label{eq:eq9}
\end{equation}
This evidential learning process is optimized by minimizing the Negative Log Likelihood ($\mathcal{L}_{\mathrm{NLL}}$) of the marginal likelihood associated with $\boldsymbol {\widetilde{x}}$. Additionally, a regularization term ($\mathcal{L}_{\mathrm{PEN}}$) is used to penalize overconfidence when evidence is weak. The detailed derivations can be found in~\cite{amini2020deep, lian2025facing}. Formally, the two loss terms are: (1) $\mathcal{L}_{\mathrm{NLL}}$, encouraging accurate modeling of reconstruction uncertainty; (2) $\mathcal{L}_{\mathrm{PEN}}$, penalizing excessive evidence for poorly reconstructed samples.
\begin{equation}
\begin{aligned}
\mathcal{L}_{\mathrm{NLL}}&=\frac12\log\left(\frac\pi v\right)-\alpha\log(\omega)+\log\left(\frac{\Gamma(\alpha)}{\Gamma\left(\alpha+\frac12\right)}\right)\\&+\left(\alpha+\frac12\right)\log\left(({\boldsymbol {x}} - \boldsymbol{\widetilde{x}})^2v+\omega\right),
\end{aligned}
\label{eq:eq10}
\end{equation}
\begin{equation}
\mathcal L_\mathrm {PEN}=| {\boldsymbol {x}} - \boldsymbol{\widetilde{x}} |\cdot(2v+\alpha),
\label{eq:eq11}
\end{equation}
where $\omega=2\beta(1+v)$ and $\Gamma(\cdot)$ is the gamma function. $\boldsymbol{\widetilde{x}}$ denotes both $\widetilde{\boldsymbol x}_{l}$ and $\widetilde{\boldsymbol x}_{h}$ multi-view branches. Notably, the higher-order parameters NIG(${v,\alpha,\beta}$) are directly predicted by a Linear layer on $\boldsymbol{\widetilde{x}}$ (As shown in the right part of~\figurename~\ref{fig:Framework}).

\noindent\textbf{Uncertainty Dynamic Fusion:}
While the low- $\widetilde{\boldsymbol x}_{l}$ and high-frequency $\widetilde{\boldsymbol x}_{h}$ multi-view branches individually capture distinct anomaly patterns, relying solely on a single frequency perspective for anomaly detection may lead to biased or incomplete results. Traditional approaches often employ score averaging or static weighting to combine information from multiple views~\cite{lian2025mfr}. However, such methods fail to adaptively account for varying levels of confidence across multi-view frequencies.

To overcome this limitation, we propose a dynamic fusion strategy that adaptively combines the evidential confidence of two frequency branches. Specifically, given the two frequency-decoupled branches, we obtain two Normal Inverse-Gamma (NIG) distributions that represent the reconstruction uncertainty: $\operatorname {NIG}(\widetilde{\boldsymbol x}_{l}, v_{l},\alpha_{l},\beta_{l})$ and $\operatorname {NIG}(\widetilde{\boldsymbol x}_{h}, v_{h},\alpha_{h},\beta_{h})$. We adopt the NIG summation mechanism proposed in~\cite{qian2018big}, which enables approximate aggregation of multiple evidential distributions to obtain fuse-view distribution $\operatorname {NIG}(\widetilde{\boldsymbol x}_{f}, v_{f},\alpha_{f},\beta_{f})$:
\begin{equation}
\begin{aligned}
\operatorname {NIG}(\widetilde{\boldsymbol x}_{f},v_{f},\alpha_{f},\beta_{f}):&=\operatorname {NIG}(\widetilde{\boldsymbol x}_{l}, v_{l},\alpha_{l},\beta_{l}) \\&\oplus \operatorname {NIG}(\widetilde{\boldsymbol x}_{h}, v_{h},\alpha_{h},\beta_{h}),
\end{aligned}
\label{eq:eq12}
\end{equation}
\begin{equation}where:\quad\widetilde{\boldsymbol x}_{f} =(v_{l}+v_{h})^{-1}(v_{l}\widetilde{\boldsymbol x}_{l}+v_{h}\widetilde{\boldsymbol x}_{h}),
\label{eq:eq13}
\end{equation}
\begin{equation}\quad v_{f}=v_{l}+v_{h},\quad\alpha_{f}=\alpha_{l}+\alpha_{h}+\frac{1}{2},
\label{eq:eq14}
\end{equation}
\begin{equation}\beta_{f}=\beta_{l}+\beta_{h}+\frac{1}{2}v_{l}(\widetilde{\boldsymbol x}_{l}-\widetilde{\boldsymbol x}_{f})^{2}+\frac{1}{2}v_{h}(\widetilde{\boldsymbol x}_{h}-\widetilde{\boldsymbol x}_{f})^{2}.
\label{eq:eq15}
\end{equation}
This approach dynamically fuses the two branches into a single unified NIG distribution. As shown in Equation~\ref{eq:eq13}, the final reconstructed value $\boldsymbol {\widetilde{x}}$ is computed as a confidence-weighted summation of the multi-view reconstructions $\widetilde{\boldsymbol x}_{l}$ and $\widetilde{\boldsymbol x}_{h}$. The view with greater confidence (i.e., $v_{l}$ or $v_{h}$) contributes more heavily to the fused distribution. And the overall uncertainty is jointly determined by: (1) The individual uncertainty of each frequency view (from its respective NIG parameters). (2) The inter-view discrepancy, i.e., the deviation between reconstructions from the low- and high-frequency branches. By dynamically summing the two evidential distributions, we obtain a new, fused NIG distribution that reflects the integrated uncertainty across both frequency perspectives.

\subsection{Training}
In the training of FreeUp, accurate reconstruction of $\widetilde{\boldsymbol x}_{l}$ and $\widetilde{\boldsymbol x}_{h}$ is a crucial basis for reliable uncertainty estimation. Poor reconstructions can collapse the estimation of marginal likelihood and lead to incorrect uncertainty quantification. Our frequency-decoupled modeling mitigates this concern by decomposing frequency components. Particularly, a $\mathcal L_\mathrm {REC}$ loss, both from spatial and frequency constraints, is jointly used to confirm accurate reconstruction. With reconstruction loss, our frequency-decoupled  framework further enhances the reliability of evidential modeling. For each frequency branch, $\widetilde{\boldsymbol x}_{l}$ and $\widetilde{\boldsymbol x}_{h}$, our loss function consists of two components, reconstruction and uncertainty estimation, and is defined as:
 \begin{equation}
\mathcal{L}= \underbrace{\mathcal{L}_{\mathrm{REC}}}_\text{Reconstruct} + \underbrace{\mathcal{L}_{\mathrm{NLL}} \cdot \mathcal{\lambda}_{\mathrm{NLL}} + \mathcal{L}_{\mathrm {PEN}} \cdot \mathcal{\lambda}_{\mathrm {PEN}}}_{\text{Uncertainty Estimate}},
\label{eq:eq16}
\end{equation}
where $\mathcal{\lambda}_{\mathrm{NLL}}$ and $\mathcal{\lambda}_{\mathrm {PEN}}$ are hyperparameters controlling the contributions of $\mathcal{L}_{\mathrm{NLL}}$ and $\mathcal{L}_{\mathrm {PEN}}$, respectively.

To further support dynamic fusion, we adopt a multi-task optimization strategy. The total loss includes not only frequency-specific objectives but also the fused loss. All components are jointly backpropagated, enabling the model to learn both frequency-specific and integrated representations effectively. The overall training objective is defined as:
\begin{equation}
\mathcal{L}^{total} = \mathcal{L}^{l} + \mathcal{L}^{h}+ \mathcal{L}^{f},
\label{eq:eq17}
\end{equation}
where the losses $\mathcal{L}^{l}$, $\mathcal{L}^{h}$ and $\mathcal{L}^{f}$ are each individually calculated using Equation~\ref{eq:eq16} for the three branches.

\subsection{Inference}
During inference, we fuse the Normal Inverse-Gamma (NIG) distribution parameters from the multi-view frequency-specific branches to form an integrated evidential representation $\operatorname {NIG}(\widetilde{\boldsymbol x}_{f}, v_{f},\alpha_{f},\beta_{f})$. This fused distribution effectively captures multi-view anomaly cues and adaptively reflects the combined confidence and uncertainty across different frequency components. To this end, we utilize the variance of the mean (i.e., epistemic uncertainty) from the fused NIG distribution as the anomaly score. This choice enables both effective and efficient anomaly detection. A higher variance indicates greater uncertainty in the model's output, suggesting that the input deviates more significantly from the learned normal patterns. This probabilistic formulation provides a principled anomaly score suitable for network traffic detection tasks. The anomaly score is computed as:
\begin{equation}
\operatorname{Anomaly \, Score} = \mathrm{Var}[\mu] = \frac{\beta_{f}}{v_{f}(\alpha_{f}-1)}.
\label{eq:eq18}
\end{equation}

\section{Experiments}\label{sec:Experiments}

\begin{table*}[t]
	\centering
	\setlength\tabcolsep{4pt}
	\belowrulesep=1pt
    \caption{\upshape Anomaly detection performance comparison on three public datasets. The best results are shown in \textbf{bold}, and the second-best are \underline{underlined}.}\label{tab:benchmark}
    \vspace{-0.1cm}
	\begin{tabular}{cccccccccc} 
	\toprule
	\multirow{2}{*}{\textbf{Methods}} & \multicolumn{3}{c}{\textbf{CIC-IoT2023}} & \multicolumn{3}{c}{\textbf{DoHBrw2020}} & \multicolumn{3}{c}{\textbf{ISCX-Tor2016}}  \\
	\cmidrule(lr){2-4} \cmidrule(lr){5-7} \cmidrule(lr){8-10}
	& \textbf{AUC(\%)}   & \textbf{ACC(\%)}  & \textbf{F1(\%)}            & \textbf{AUC(\%)}   & \textbf{ACC(\%)}  & \textbf{F1(\%)}              & \textbf{AUC(\%)}   & \textbf{ACC(\%)}  & \textbf{F1(\%)}             \\ 
	\toprule	
DAGMM               & 66.40 & 61.38 & 67.22           & 77.41 & 75.18 & 75.07          & 73.28 & 74.30 & 73.82            \\
GANomaly            & 67.43 & 65.74 & 72.02           & 79.34 & 79.87 & 79.83          & 92.42 & 89.64 & 89.77            \\
ARCADE              & 72.09 & 77.32 & 74.94           & 76.79 & 79.91 & 80.07          & 92.35 & 90.15 & 90.50            \\
MFR                 & 82.68 &  \underline{83.03} & 81.56           & 70.51 & 70.23 & 73.44          & 93.68 & 91.74 & \underline{91.94}           \\
UnDiff              &  \underline{83.25} & 82.69 & \underline{81.93}         & \underline{81.92} & 79.93 & 80.10   & \underline{94.01} & \underline{91.86} & 91.61           \\
\midrule
NeuTral             & 81.12 & 80.49 & 81.15           & 72.44 & 70.98 & 76.55          & 92.06 & 90.82 & 90.70            \\
MCM        & 68.99 & 71.77 &  73.38           &  80.86 & \underline{82.55} & \underline{81.87}         & 93.01 & 91.11 & 91.26            \\
Anomaly Transformer & 64.63 & 67.15 & 71.86           & 78.86 & 78.55 & 78.87          & 92.27 & 90.19 & 90.52     \\
TSLANet             & 81.64 & 76.62 & 80.04           & 77.04 & 68.93 & 73.10          & 93.63 & 90.97 & 90.71      \\
\midrule
\textbf{FreeUp}             & \textbf{86.68} & \textbf{84.67} & \textbf{85.55}           &   \textbf{85.44}    &    \textbf{84.96}   &   \textbf{84.32}    & \textbf{95.53} & \textbf{93.26} & \textbf{93.22}        \\
	\bottomrule
	\end{tabular}
\label{Table:MainResults}
\end{table*}

\subsection{Experimental Setups}
\noindent\textbf{Dataset.} We evaluate our approach using three publicly available encrypted traffic datasets representing diverse anomaly scenarios. (1) CIC-IoT2023~\cite{neto2023ciciot2023} is a encrypted intrusion detection dataset specifically designed for Internet of Things (IoT) network environments. (2) DoHBrw2020~\cite{montazerishatoori2020detection} focuses on detecting malicious traffic within HTTPS communications. (3) ISCX-Tor2016~\cite{lashkari2017characterization} contains anonymous network traffic commonly employed to conceal malicious activities, making it particularly suitable for evaluating detection capabilities against stealthy anomalies~\cite{yuan2024phogad, zheng2023semi}. Following established experimental protocols~\cite{lian2025facing}, we construct our training set by randomly sampling 10,000 instances of normal traffic. For evaluation, we create a balanced test set comprising 5,000 normal and 5,000 anomalous traffic instances. 

\noindent\textbf{Evaluation Metrics.} Following established practices in network anomaly detection~\cite{lian2025facing, lian2025mfr}, we employ three standard metrics to evaluate detection performance: AUROC (AUC), Accuracy (ACC), F1-Score (F1).

\noindent\textbf{Baselines.} To comprehensively evaluate the effectiveness of FreeUp, we conduct comparative experiments against nine state-of-the-art baselines spanning two categories: (1) {Network traffic anomaly detection}: DAGMM~\cite{zong2018deep}, GANomaly~\cite{akcay2018ganomaly}, ARCADE~\cite{lunardi2023arcade}, MFR~\cite{lian2025mfr}, and UnDiff~\cite{lian2025facing}. (2) {Time series anomaly detection}:  NeuTral~\cite{qiu2021neural}, MCM~\cite{yin2024mcm}, Anomaly Transformer~\cite{xu2022anomaly}, and TSLANet~\cite{eldele2024tslanet}.

\noindent\textbf{Implementation Details.} All experiments are conducted on an NVIDIA GeForce RTX 4090 GPU. We use the Adam optimizer with a $1\times10^{-3}$ learning rate and a batch size of 128. The maximum training epoch is 60, with early stopping to prevent overfitting. Following~\cite{lian2025facing}, we set the uncertainty loss coefficients as $\mathcal{\lambda}_{\mathrm{NLL}}=1\times10^{-2}$ and $\mathcal{\lambda}_{\mathrm {PEN}}=1\times10^{-4}$ across all three datasets. We also fix $\mathcal{\lambda}_{\mathrm{F}}=1\times10^{-2}$ to promote effective frequency-domain learning. For hyperparameters, we set the number of packets $P$ in a traffic sample to 8, and the Gaussian filter cutoff frequency $D$ is 5. To ensure statistical robustness, we report the mean results over five independent runs with different random seeds.

\begin{figure}[t]
\centerline{\includegraphics{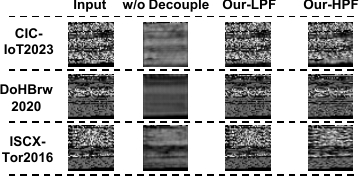}}
\vspace{-0.18cm}
\caption{Reconstruction comparisons before and after frequency decoupling.}
\label{fig:Recon}
\end{figure}
\subsection{Main Results}
Table~\ref{Table:MainResults} compares FreeUp and baselines. Based on these results, we make the following observations:

\noindent\textbf{(O1)}: Our FreeUp consistently outperforms all baselines across various datasets and evaluation metrics by substantial margins. Specifically, FreeUp achieves AUC improvements of over 3\% on the CIC-IoT2023 and DoHBrw2020 datasets and up to 95\% AUC performance in ISCX-Tor2016. These results validate the effectiveness of the frequency-decoupled framework. The frequency-decoupled manner not only facilitates complete pattern understanding for ``full-frequency'' traffic image but also provides the multi-view evidence modeling for uncertainty-inspired evidential learning. Based on the multi-view anomaly criteria, our evidential learning fusion method further adaptively integrates the anomaly evaluations from different frequency perspectives, leading to a more comprehensive and accurate anomaly detection.

\noindent\textbf{(O2)}: Majority of network traffic anomaly detection methods are based on an image-format process manner. However, the ``full-frequency'' traffic images are deeply disturbed by ``spectral mismatch'' issue in neural networks, which cannot effectively construct the profile of normal traffic during the training (as shown by our analytical and experimental proofs in Section~\ref{sec:Motivation}). It further limits the reliable distinction of anomalous traffic during testing. GANomaly and ARCADE methods do not produce promising results across three datasets. MFR and UnDiff methods focus only on low-frequency texture features in traffic images to alleviate this problem by imposing the low-pass filters. However, they discard the role of high-frequency information entirely and use incomplete information to carve out anomalies, causing unreliable detection results.

\noindent\textbf{(O3)}: Time-series-based methods, like Anomaly Transformer and TSLANet, mainly emphasize the transformer model's ability to capture the trends and fluctuations anomalies in sequential data. MCM captures intrinsic correlations between feature data. NeuTral is a contrastive learning-based spatial transformation method for tabular data. However, due to encryption, network traffic often loses these original sequential and tabular characteristics, negatively impacting the effectiveness of these time-series-based methods' anomaly criteria.

\begin{figure}[t]
\centerline{\includegraphics{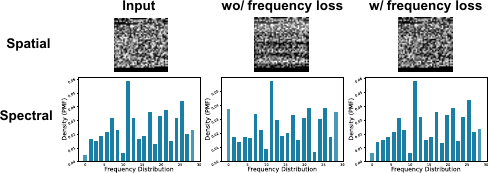}}
\vspace{-0.18cm}
\caption{Spatial–spectral comparison under the frequency-constrained loss.}
\label{fig:FrequencyLoss}
\end{figure}

\begin{table}[t]
\setlength\tabcolsep{3pt}
\belowrulesep=0.8pt
\centering
\caption{\upshape Ablation study on frequency decoupling using AUC metric.}
\vspace{-0.1cm}
\begin{tabular}{ccccc}
\toprule
 \textbf{Variant} & \begin{tabular}[c]{@{}c@{}}\textbf{CIC-IoT2023}\end{tabular} & \begin{tabular}[c]{@{}c@{}}\textbf{DoHBrw2020}\end{tabular} & \begin{tabular}[c]{@{}c@{}}\textbf{ISCX-Tor2016}\end{tabular}  \\
\toprule
w/o HB   &   77.09     &  83.62     &    95.10     \\
w/o LB   &   68.73     &  82.73     &    94.37     \\
w/o FD   &   82.10     &  81.26     &    92.80     \\
w/o FC   &  85.57      &  83.85     &    94.62    \\
\midrule
\textbf{FreeUp}     & \textbf{86.68}       & \textbf{85.44}      & \textbf{95.53}       \\
\bottomrule
\end{tabular}
\label{Table:fre_ablation}
\end{table}

\begin{figure*}[t]
\centerline{\includegraphics{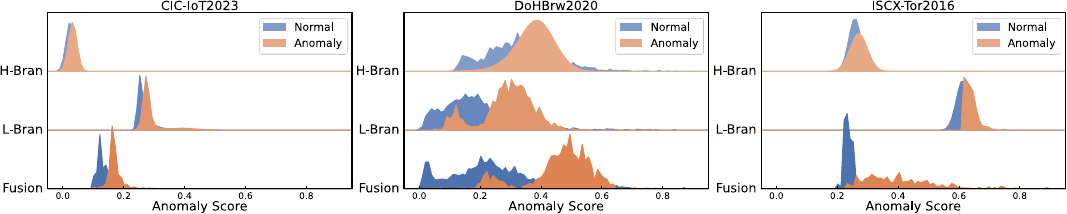}}
\vspace{-0.18cm}
\caption{Dynamic fusion of anomaly scores (density distribution) after multi-task learning across three datasets.}
\label{fig:Fusion}
\end{figure*}
\subsection{Frequency Decoupling Study}

\noindent\textbf{\textit{Ablation Experiments}}: We conduct ablation studies to evaluate the effectiveness of frequency decoupling, as shown in Table~\ref{Table:fre_ablation}. Variants include removing the low-frequency branch (w/o LB), high-frequency branch (w/o HB), the frequency-decoupled framework (w/o FD), and the frequency-constrained loss (w/o FC). The single-branch training leads to a noticeable drop in detection performance. This suggests the incomplete frequency modeling introduces biased anomaly evaluation, especially in the absence of the low-frequency branch (w/o LB). Removing the frequency-decoupled framework (w/o FD) results in the most significant degradation, highlighting its critical role in FreeUp. This degradation is attributed to the ``full-frequency'' phenomenon, which impairs traffic image reconstruction and subsequent uncertainty modeling. Moreover, excluding the frequency-constraint loss (w/o FC) also reduces performance, verifying its effectiveness in guiding each branch's frequency-specific learning.

\noindent\textbf{\textit{Qualitative Experiments}}: To further verify the benefits of frequency decoupling, we visualize reconstruction results, as shown in~\figurename~\ref{fig:Recon}. Due to the ``full-frequency'' nature of network traffic, models without frequency decoupling (w/o Decouple) consistently exhibit poor reconstruction, like the same empirical validation in Section~\ref{sec:Motivation}. This further supports the presence of a ``spectral mismatch'' issue, wherein inadequate modeling results in limited pattern understanding. In contrast, FreeUp, which explicitly decouples and processes frequency components via dedicated branches, achieves more faithful reconstructions and thus more accurate pattern modeling. This improvement arises from the model's ability to focus on localized frequency bands, freeing up the burden of full-spectrum representation. Furthermore, the complementary integration of low- and high-frequency branches enhances reconstruction stability and enables richer, multi-view anomaly quantification. Overall, frequency decoupling alleviates the model's spectral bias and facilitates a more complete representation of traffic distributions. The spatial and spectral visualization of frequency loss in~\figurename~\ref{fig:FrequencyLoss} further supports the effectiveness of stable frequency learning and demonstrates the benefits introduced by the frequency-constrained loss.

\begin{table}[t]
\setlength\tabcolsep{2.8pt}
\belowrulesep=0.8pt
\centering
\caption{\upshape Ablation study on dynamic fusion using AUC metric.}
\vspace{-0.1cm}
\begin{tabular}{ccccc}
\toprule
 \textbf{Variant} & \begin{tabular}[c]{@{}c@{}}\textbf{CIC-IoT2023}\end{tabular} & \begin{tabular}[c]{@{}c@{}}\textbf{DoHBrw2020}\end{tabular} & \begin{tabular}[c]{@{}c@{}}\textbf{ISCX-Tor2016}\end{tabular}  \\
\toprule
Low-Branch &   75.57     &   77.53    &    92.33     \\
High-Branch &   57.24     &   67.30    &    78.27   \\
Product-Fuse &   74.76     &   75.39    &    90.47     \\
Sum-Fuse &   75.56     &   77.29    &    92.34     \\
\midrule
\textbf{FreeUp}     & \textbf{86.68}       & \textbf{85.44}      & \textbf{95.53}       \\
\bottomrule
\end{tabular}
\label{Table:fusion_ablation}
\end{table}

\subsection{Dynamic Fusion Study}
\noindent\textbf{\textit{Ablation Experiments}}: To evaluate the effectiveness of our dynamic uncertainty fusion strategy, we compare it with common anomaly score fusion schemes, score product (Product-Fuse) and weighted summation (Sum-Fuse)~\cite{lian2025mfr}, based on the jointly trained multi-branch architecture (Low-Branch and High-Branch). As shown in Table~\ref{Table:fusion_ablation}, conventional fusion methods consistently fail to improve performance in some cases, and even lead to performance degradation. This is primarily due to their reliance on static weighting strategies, which cannot adaptively account for the varying contributions of each branch. In contrast, our dynamic uncertainty fusion method inherently leverages the distributional parameters from multiple frequency branches, enabling a more flexible and comprehensive integration of anomaly signals.

\noindent\textbf{\textit{Qualitative Experiments}}:
\figurename~\ref{fig:Fusion} reveals our fusion process from multiple frequency perspectives across the three datasets. Although the jointly trained multi-view branches in Table~\ref{Table:fusion_ablation} shows slight performance degradation compared to individually trained branches in Table~\ref{Table:fre_ablation}, our method still enables dynamic fusion of anomaly score (i.e., uncertainty distribution) and adaptively integrates jointly trained multi-branch information. Like the anomaly score distribution of high-frequency branch (H-Bran) in~\figurename~\ref{fig:Fusion}, they exhibit the most confused anomaly score distributions (overlapping the blue and yellow distribution). However, through dynamic fusion, these branches still contribute proper complementary signals that are integrated with those from the low-frequency branch (L-Bran), ultimately producing a more distinguishable fusion anomaly score. These results validate both the effectiveness of our multi-task training strategy and the success of the proposed adaptive fusion mechanism in capturing nuanced multi-frequency anomaly patterns.

\subsection{Hyperparameter Study}

\begin{figure}[t]
\centerline{\includegraphics{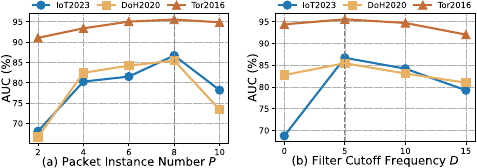}}
\vspace{-0.18cm}
\caption{Sensitivity analysis of key hyperparameters across three datasets.}
\label{fig:Hyper}
\end{figure}

\figurename~\ref{fig:Hyper} presents the sensitivity analysis of two key hyperparameters in FreeUp. The observations are as follows:

\noindent\textbf{(O1)}: The small number of packet instances $P$ may lead to insufficient contextual information for reliable anomaly scoring, whereas an excessively large $P$ introduces noise and may obscure critical patterns. Empirically, setting $P=8$ achieves the best trade-off, consistent with previous findings~\cite{lian2025mfr}.

\noindent\textbf{(O2)}: The parameter $D$ governs the bandwidth radius of the Gaussian filters used to decouple low- and high-frequency components. It determines the extent of separation between frequency branches, which is crucial for frequency-sensitive uncertainty modeling and effective multi-view anomaly detection. The results indicate that $D=5$ provides the most suitable frequency decoupling extent.

\begin{table}
\centering
\setlength\tabcolsep{15pt}  
\belowrulesep=0.8pt
\caption{\upshape Overhead comparison for inference.}\label{tab:overhead}
\vspace{-0.1cm}
\begin{tabular}{ccc}
\toprule
\textbf{Methods}            & \textbf{\#MACs (G)} & \textbf{\#Paras (M)}  \\
\toprule
ARCADE             & 0.82     & 6.70          \\
MFR                & 1.01     & 11.18        \\
UnDiff             & \textbf{0.25}     & \textbf{2.55}         \\
NeuTral            & 1.45     & 139.57       \\
MCM        & 0.84     & 822.53         \\
TSLANet            & 6.45     & 179.19       \\
\midrule
\textbf{FreeUp}       & \underline{0.73}     &  \underline{6.46}        \\
\bottomrule
\end{tabular}
\end{table}
\subsection{Overhead Analysis}
Table~\ref{tab:overhead} summarizes the overhead comparison of the most competitive approaches in terms of multiply-accumulate operations per second (\#MACs) and the number of model parameters (\#Paras). Overall, our method, FreeUp, demonstrates promising efficiency compared to detection approaches such as MFR and ARCADE. This is attributed to our concise autoencoder design and lightweight evidential learning framework. Transformer-based models like TSLANet incur high computational costs due to attention operations, while NeuTral and MCM require multiple models integration, increasing parameter overhead. Although FreeUp’s dual-branch design results in roughly three times the overhead of UnDiff, this cost is offset by superior detection performance. Moreover, thanks to our lightweight structure, the dual branches can be deployed in parallel with minimal memory usage, further improving practical runtime efficiency.
\section{Related Work}\label{sec:Related Work}

Many studies transform raw traffic data into image-like representations and employ feature extraction techniques to improve effectiveness. MFAD~\cite{zheng2023semi} and MFR~\cite{lian2025mfr} first observed the non-textured and irregular spatial characteristics of traffic images. They applied low-pass filtering to emphasize low-frequency texture-level information. Similarly, unFlowS~\cite{yang2025unflows} maps complex traffic data into a simplified spectral space and leverages statistical properties for anomaly detection. To enhance representation learning, ARCADE~\cite{lunardi2023arcade} incorporate adversarial training mechanisms to improve the fitting of normal data distributions. More recently, UnDiff~\cite{lian2025facing} reveals the intra-sample ``identical shortcut'' limitation in disordered traffic images and proposes an inter-sample comparison framework that utilizes filtered distributional differences for anomaly quantification. While these methods acknowledge the spatial domain difficulties posed by ``full-frequency'' traffic data, most do not explicitly address how to model frequency-specific patterns. In fact, the bias toward low-frequency features in full-spectrum modeling often hinders effective representation learning and anomaly identification.

Since network traffic can also be represented as time series data~\cite{lian2024payload, dang2025semi},  we also consider time-series-based anomaly detection approaches, particularly Transformer-based architectures, for comprehensive evaluation. These methods typically model sequential dependencies, such as periodic trends and temporal fluctuations. For instance, Anomaly Transformer~\cite{xu2022anomaly} introduces an anomaly-attention mechanism to capture both prior associations and series-wise dependencies within temporal sequences. TimesNet~\cite{wu2023timesnet} addresses period variation challenges by employing stack inception blocks to unify multi-scale temporal patterns. TSLANet~\cite{eldele2024tslanet} integrates the Fourier Transform with self-supervised learning techniques to model periodic dependencies. NeuTral~\cite{qiu2021neural} transforms semantic space to comparative learning spaces for anomaly assessment, while MCM~\cite{yin2024mcm} employs a masking strategy to learn intrinsic correlations. Despite the demonstrated effectiveness of these models on clean time-series datasets, their applicability to network traffic anomaly detection remains problematic. The ``full-frequency'' phenomenon observed in traffic images also exists in sequential formats. The high-frequency component can obscure underlying trends and fluctuations in sequential data, similar to how it suppresses texture features in images. This significantly degrades the core temporal dependencies necessary for effective anomaly detection.

\section{Conclusion}\label{sec:Conclusions}
In this paper, we are the first to identify the pervasive ``full-frequency'' phenomenon in network traffic data, which poses a significant ``spectral mismatch'' challenge for network traffic analysis. To address this, we propose FreeUp, a novel frequency-decoupled framework that effectively decomposes complex ``full-frequency'' information for improved understanding and generates complementary multi-view representations. We further design a parameterized dynamic uncertainty fusion mechanism to adaptively integrate multi-view frequency anomaly cues. Extensive quantitative and qualitative experiments consistently demonstrate the effectiveness of FreeUp on multiple real-world traffic datasets.

\section*{Acknowledgment}
This work was supported by National Natural Science Foundation of China (Grant No. 62572097 and No. 62176043).

\bibliographystyle{unsrt}
\bibliography{infocom}

\end{document}